\documentclass[conference]{IEEEtran}
\IEEEoverridecommandlockouts

\usepackage{cite}
\usepackage{amsmath,amssymb,amsfonts}
\usepackage{multirow}
\usepackage{algorithmic}
\usepackage{graphicx}
\usepackage{textcomp}
\usepackage{xcolor}

\usepackage{booktabs}
\usepackage{float}
\usepackage{multirow}
\usepackage{makecell}
\usepackage{bm}
\usepackage{xspace}
\usepackage{footnote}
\makeatletter
\DeclareRobustCommand\onedot{\futurelet\@let@token\@onedot}
\def\@onedot{\ifx\@let@token.\else.\null\fi\xspace}

\def\ie{\emph{i.e}\onedot}

\def\etal{\emph{et al}\onedot}
\makeatother
\usepackage{epstopdf}

\def\BibTeX{{\rm B\kern-.05em{\sc i\kern-.025em b}\kern-.08em
T\kern-.1667em\lower.7ex\hbox{E}\kern-.125emX}}

\begin{document}

\title{Temporal-Frequency State Space Duality: An Efficient Paradigm for Speech Emotion Recognition}

\author{
\IEEEauthorblockN{Jiaqi Zhao$^{1,2}$, Fei Wang$^{2,3,6*}$, Kun Li$^5$, Yanyan Wei$^3$, Shengeng Tang$^3$, Shu Zhao$^4$, Xiao Sun$^{2,3}$}
\IEEEauthorblockA{$^1$ School of Artificial Intelligence, Anhui University, Hefei, China}
\IEEEauthorblockA{$^2$ Institute of Artificial Intelligence, Hefei Comprehensive National Science Center, Hefei, China}
\IEEEauthorblockA{$^3$ School of Computer Science and Information Engineering, Hefei University of Technology, Hefei, China}
\IEEEauthorblockA{$^4$ School of Computer Science and Technology, Anhui University, Hefei, China}
\IEEEauthorblockA{$^5$ CCAI, Zhejiang University, Zhejiang, China~~$^6$ Hefei Zhongjuyuan Intelligent Technology Co., Ltd., Hefei, China}
\thanks{* Corresponding author}
}
\maketitle

\begin{abstract}
Speech Emotion Recognition (SER) plays a critical role in enhancing user experience within human-computer interaction. 
However, existing methods are overwhelmed by temporal domain analysis, overlooking the valuable envelope structures of the frequency domain that are equally important for robust emotion recognition. 
To overcome this limitation, we propose TF-Mamba, a novel multi-domain framework that captures emotional expressions in both temporal and frequency dimensions.
Concretely, we propose a temporal-frequency mamba block to extract temporal- and frequency-aware emotional features, achieving an optimal balance between computational efficiency and model expressiveness. 
Besides, we design a Complex Metric-Distance Triplet (CMDT) loss to enable the model to capture representative emotional clues for SER. Extensive experiments on the IEMOCAP and MELD datasets show that TF-Mamba surpasses existing methods in terms of model size and latency, providing a more practical solution for future SER applications.

\end{abstract}


\section{Introduction}
Understanding human emotions through speech~\cite{shen2024asr,chen2023dwformer,chen2023dst,chen2023speechformer++,zhou2024audio} has emerged as a critical frontier in social interactions~\cite{chen2024prototype,li2023data,li2024patch,qia2024cluster}, with the potential to revolutionize applications ranging from mental health monitoring to adaptive user interfaces.
Speech Emotion Recognition (SER), which aims to identify emotional states from vocal cues, presents a significant challenge due to the inherent variability in speech patterns and the subtlety of emotional expressions~\cite{chen2023dst,chen2023dwformer}.





Previous works have primarily focused on fusing acoustic features across modalities to enrich emotional expression.
For instance, 
Zou~\etal~\cite{zou2022speech} combined hand-created acoustic features with deep learning-based features for co-learning. However, this approach may limit the autonomous learning potential of high-dimensional complex features~\cite{chen2022wavlm,tang2021graph,tang2024gloss}, as it can introduce inter-feature conflicts or redundancies.
Shen~\etal~\cite{shen2024asr} introduced Automatic Speech Recognition (ASR) to model more fine-grained emotional features.
On the other hand, while improvements of the attention mechanism inside Transformer architectures\cite{chen2022speechformer,chen2023speechformer++,chen2023dwformer,chen2023dst,shen2023mingling,shen2023adaptive,shen2023mutual,wei2021deraincyclegan,wang2024low} have been shown to improve SER accuracy and generalization across multiple languages by capturing saliency information, these models tend to focus only on intensity variations in the temporal domain. 
Besides, the stacking of Transformer blocks often prevents the balance between accuracy and computational efficiency, limiting their practicality.
\begin{figure}[t!]
\begin{center}
\centerline{\includegraphics[width=1\linewidth]{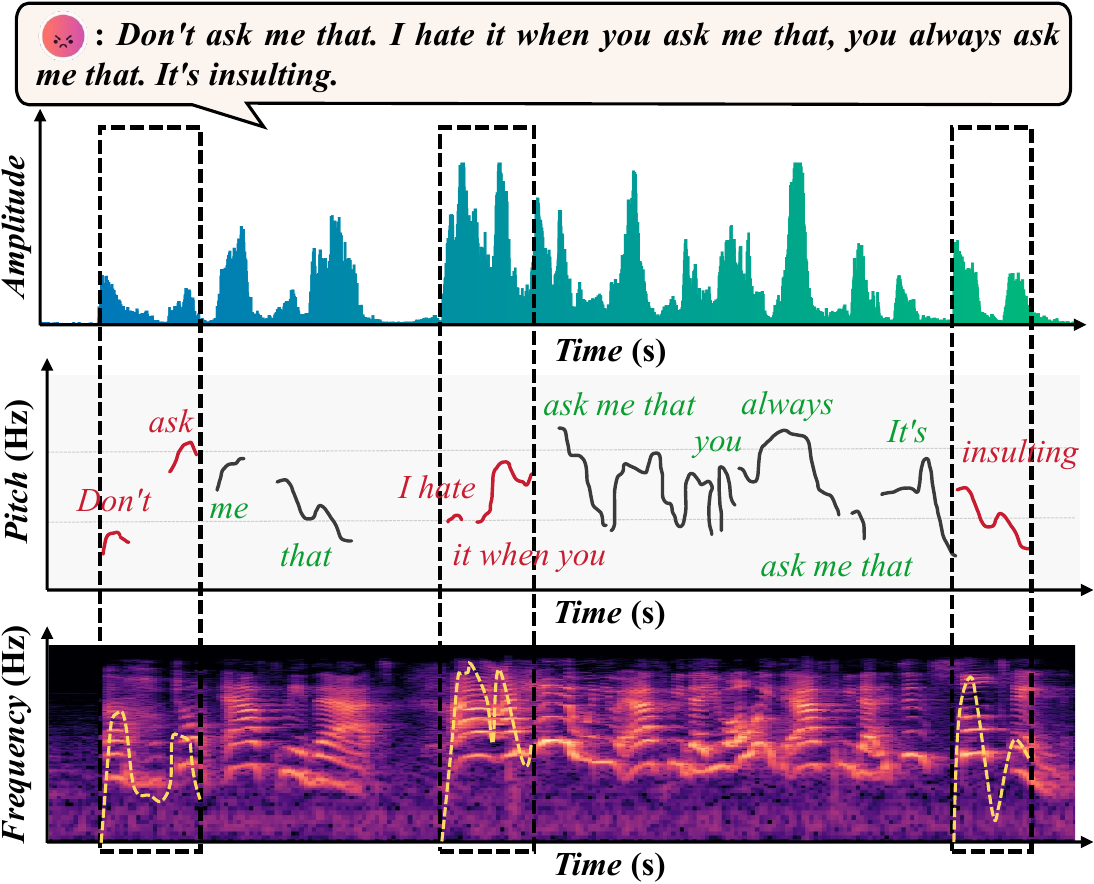}}
\caption{Visualization of time-aligned speech intensity, pitch, and frequency spectrum for a real sample from the IEMOCAP dataset~\cite{busso2008iemocap}.
The frequency spectrum reveals finer structures of tone, timbre, and other envelope structures.
}
\label{fig:bg}
\end{center}
\vspace{-2 em}
\end{figure}

To tackle the challenge, we focus on two key aspects:
(a) Given the automatic learning capability of complex speech features~\cite{chen2022wavlm}, we mainly focus on their high-dimensional frequency distribution. 
Inspired by \cite{gfeller2020spice}, Fig.\ref{fig:bg} visualizes a real speech signal's intensity, pitch, and spectrum aligned along time. The frequency energy space provides more finely detailed frequency envelope structures, such as tone and timbre, to capture richer emotional information.
(b) The State Space Duality (SSD)~\cite{mamba,mamba2,liu2024micro} of the Mamba architecture is naturally suited for processing long temporal sequences. It excels at capturing the continuous evolution of acoustic features. Besides, Mamba offers significant computational complexity and efficiency advantages compared to traditional Transformer structures~\cite{chen2023dwformer,chen2023speechformer++,chen2023dst,wang2024eulermormer,hu2024maskable,li2023multi}.
\begin{figure*}[t!]
\begin{center}
\includegraphics[width=1\linewidth]{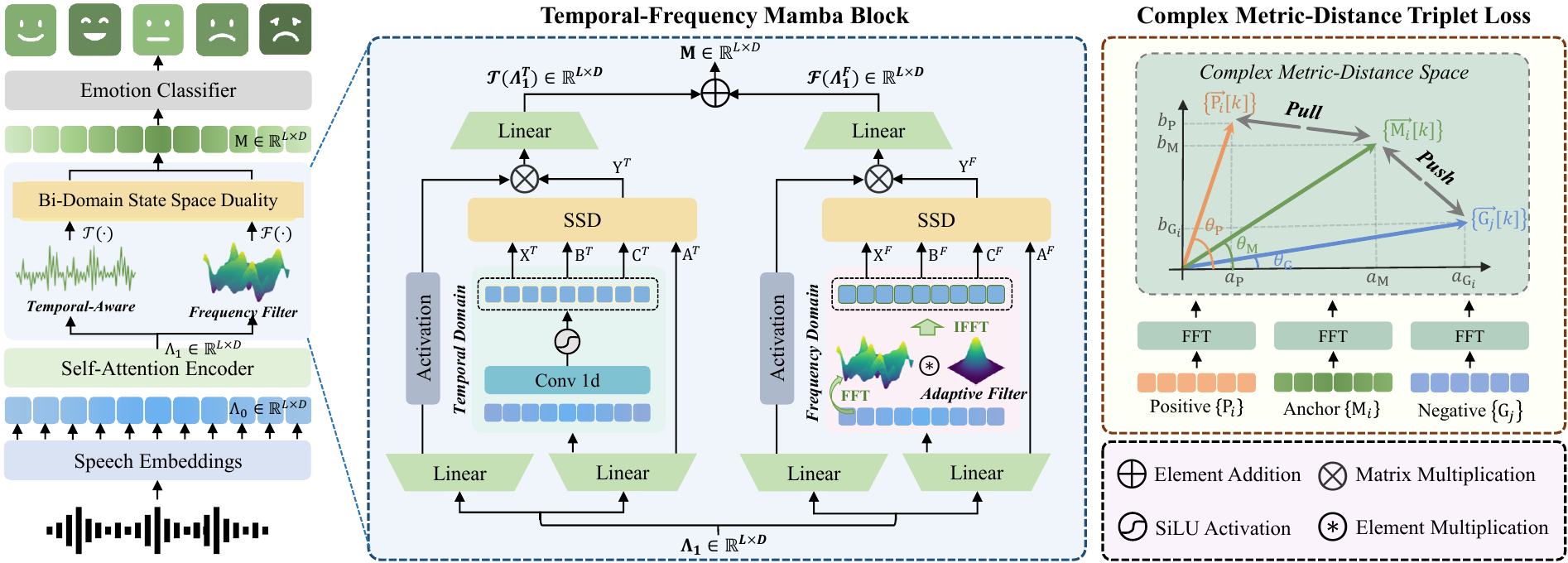}
\vspace{-1.5 em}
\caption{The proposed TF-Mamba framework introduces an innovative multi-domain learning paradigm designed to precisely capture speech emotion expressions in both temporal and frequency domains. TF-Mamba optimizes computational efficiency and model performance by leveraging an efficient Bi-Domain SSD mechanism and integrating temporal perception and frequency filtering modules. Besides, the CMDT loss enhances the clustering of emotional samples and the separation of emotions, improving emotional discrimination and model robustness.}
\label{fig:overall}
\end{center}
\vspace{-1.5 em}
\end{figure*}
Based on these considerations, our main contributions are as follows:
\begin{itemize}
\item We propose a novel multi-domain learning paradigm for SER, centered on a \textbf{T}emporal-\textbf{F}requency \textbf{Mamba} (\textbf{TF-Mamba}) designed to capture the combined temporal and frequency-based speech emotional expressions precisely.
\item TF-Mamba leverages temporal awareness and frequency filtering within an efficient Bi-Domain SSD, optimizing the balance between computational efficiency and model expressiveness for high-dimensional speech modeling.
\item We introduce an innovative Complex Metric-Distance Triplet (CMDT) loss, which enhances the clustering of similar emotional samples and the separation of different emotions in fine-grained utterance-level SER, improving the model’s emotional discrimination and robustness.
\item Extensive experiments show that we outperform existing methods with fewer parameters and faster latency.
\end{itemize}

\section{Proposed Method}


\subsection{Overview}\label{sec:overview}
We aim to design an efficient multi-domain learning paradigm to address the limitations of traditional acoustic temporal features in SER tasks. As shown in Fig.\ref{fig:overall}, the proposed TF-Mamba first processes the speech signal through WavLM-Large to extract embeddings $\Lambda_0\!\in\!\mathbb{R}^{L\times D}$, refined through a multi-head self-attention mechanism to obtain shallow features $\Lambda_1\!\in\!\mathbb{R}^{L\times D}$, where $L$ denotes the number of feature tokens and $D$ is the feature dimension.
Next, the core TF-Mamba block performs deep feature extraction via State Space Duality in both the temporal and frequency domains, generating complex and interconnected speech emotion features $\operatorname{M}\!\in\!\mathbb{R}^{L\times D}$. Finally, a multi-layer perceptual emotion classifier predicts emotions with precision. Notably, we introduce a CMDT loss to enhance the clustering of similar emotional samples and the separation of distinct ones in fine-grained utterance-level SER.

\subsection{Temporal-Frequency Mamba Block}
Effectively capturing subtle emotional expressions in complex speech features is crucial. Therefore, we leverage the advanced SSD mechanism in the Mamba architecture and propose a Bi-Domain SSD structure to construct the Temporal-Frequency Mamba block, enabling precise temporal and frequency-based comprehensive speech emotion representation.
Specifically, the normalized input $\Lambda_{1}$ is fed into the temporal-domain mamba branch $\mathcal{T}(\cdot)$ and the frequency-domain mamba branch $\mathcal{F}(\cdot)$ to model both the temporal- and frequency-aware features. 
The resulting outputs, $\mathcal{T}(\Lambda_1^T)$ and $\mathcal{F}(\Lambda_1^F)$, are then fused via concatenation $\operatorname{Concat}(\cdot)$ and connected to the residual to obtain temporal and frequency-based speech emotion representation $\operatorname{M}$.
The complete process can be summarized as:
\begin{equation}
\begin{aligned}
\operatorname{M}=  \Lambda_1 + \operatorname{Concat}(\mathcal{T}(\Lambda_1^T),~\mathcal{F}(\Lambda_1^F))\!\in\!\mathbb{R}^{L\times D}.
\end{aligned}
\end{equation}

Notably, $\mathcal{T}(\cdot)$ and $\mathcal{F}(\cdot)$ are designed based on the Mamba blocks of the Bi-Domain SSD structure, \ie, for the linear projections
$\{\Phi^{T}\!\in\!\mathbb{R}^{L\times(2D'+2G\cdot d)},\Phi^{F}\!\in\!\mathbb{R}^{L\times(2D'+2G\cdot d)}\}$ 
of inputs $\{\Lambda^{T},\Lambda^{F}\}$, where $D'$ and $d$ denote the expansion and state dimension and $G$ means the number of groups, the temporal and frequency-domain SSD structures leverage their parallel duality properties to model deep and complex information across different domains efficiently. 

\subsubsection{\bfseries Temporal-Aware Module}
To understand how temporal information in speech signals is used for SSD modeling, we apply a 1D Conv layer $W_{1d}(\cdot)$ to the high-dimensional projection $\Phi^{T}$, sharing convolutional kernel weights along the time dimension to capture short-term dependencies in speech. Then, the SiLU activation function $\sigma$ dynamically adjusts the weights of the inputs, allowing for more precise handling of subtle emotional variations across different time steps. Consequently, the state vector $\mathbf{X}^{T}\!\in\!\mathbb{R}^{L\times D'}$, input matrix $\mathbf{B}^{T}\!\in\!\mathbb{R}^{L\times(G\cdot d)}$, output matrix $\mathbf{C}^{T}\!\in\!\mathbb{R}^{L\times(G\cdot d)}$, and state transition matrix $\mathbf{A}^{T}\!\in\!\mathbb{R}^{L\times D'}$ modeled by the temporal-aware module as:
\begin{equation} \label{ssd}
\begin{aligned}
\{ \mathbf{X}^{T},\mathbf{B}^{T},\mathbf{C}^{T},\mathbf{A}^{T}\} = \mathrm{Split}(\sigma \cdot (W_{1d}(\Phi^{T}))).
\end{aligned}
\end{equation}

\subsubsection{\bfseries Frequency Filter Module}
Inspired by \cite{rao2021global,singh2021non,li2023speech}, the low-frequency components of speech signals contain a significant amount of fundamental frequency information, while emotional states are often primarily reflected in pitch and intonation changes as reflected by the low-frequency region.
Therefore, as for $\{\Phi^{F}_0\!\in\!\mathbb{R}^{L\times(D'+2G\cdot d)}$$,$$\mathbf{A}^{F}\!\in\!\mathbb{R}^{L\times D'}\}$ = $\mathrm{Split}(\Phi^{F})$, it is first transformed into the frequency domain along $L$ via the 1D Fast Fourier Transform $\mathrm{FFT}(\cdot)$ as:
\begin{equation}\label{fft}
\begin{aligned}
\Theta^{F}= \mathrm{FFT}(\Phi^{F}_0)\!\in\!\mathbb{R}^{L'\times(D'+2G\cdot d)},
\end{aligned}
\end{equation}
where $L'$ is the transformed sequence length of the frequency domain representation $\Theta^{F}$.
To effectively disentangle the fundamental frequency information, we designed an adaptive low-pass filter that can learn appropriate frequency thresholds $\omega$ from $\Theta^{F}$ to optimize the capture of emotional features. Subsequently, the filtered parameter matrices $\{\mathbf{X}^{F}\!\in\!\mathbb{R}^{L\times D'},\mathbf{B}^{F}\!\in\!\mathbb{R}^{L\times(G\cdot d)},\mathbf{C}^{F}\!\in\!\mathbb{R}^{L\times(G\cdot d)}\}$ are reconstructed by inverse Fourier transform $\mathrm{IFFT}(\cdot)$, generating features with enhanced emotional expressiveness.
\begin{equation}\label{ifft}
\begin{aligned}
\{\mathbf{X}^{F},\mathbf{B}^{F},\mathbf{C}^{F}\}= \mathrm{Split}(\mathrm{IFFT}(\Theta^{F}\odot(\left|\Theta^{F}\right|^2 >\omega)),
\end{aligned}
\end{equation}
where $\left|\Theta^{F}\right|^2$ is the power spectrum, measuring the intensity of different frequencies within the speech features.

\subsubsection{\bfseries Bi-Domain State Space Dualit (SSD)}
SSD, as an efficient enhancement to the basic State Space Model (SSM)~\cite{mamba}, introduces a structured duality mask matrix to describe the evolution of the state space, allowing parallel processing across different time steps to efficiently derive the mapping relationship $y \leftarrow \{x, B, C, A\}$.
Both SSD and SSM are derived from the system of equations in Eq.~\ref{ssd0}.
\begin{equation}\label{ssd0}
\begin{aligned}
h_{t}&=A_th_{t-1}+B_tx_t,\\
y_{t}&=C_t^\top h_t,
\end{aligned}
\end{equation}
where $x_t$ and $y_t$ are treated as scalars, denoting the input and output at time $t$, respectively. $h_t$ is the hidden state vector, and the three matrix parameters ($A_t$, $B_t$, $C_t$) can vary over $t$.


The SSD in Mamba2~\cite{mamba2} simplifies the diagonal matrix $A_t$ by setting all diagonal elements to the same value and all off-diagonal elements to zero, which increases computing performance. Thanks to the large reduction in computing cost, the attention mechanism can be integrated with SSD while retaining the model's structural integrity.
Specifically, SSD employs a single projection at the block's entry point to simultaneously handle the state vector \( X \), state transition matrix \( A \), input matrix \( B \), and output matrix \( C \), modeling the mapping from \([X, A, B, C]\) to \( Y \). This mechanism is analogous to how standard attention architectures generate \( Q \), \( K \), and \( V \) projections in a parallel manner.
Therefore, the SSD mechanism is applied to both the temporal-aware and frequency filter, enabling precise extraction of complex emotional features with lower computational cost, thereby improving the overall SER performance.



Combining Eqs.\ref{ssd} and \ref{ssd0}, we obtain the outputs $\{\mathbf{Y}^{T},\mathbf{Y}^{F}\}$ as:
\begin{equation}
\begin{aligned}
\mathbf{Y}^{T}&=\mathrm{SSD}(\mathbf{A}^{T},\mathbf{B}^{T},\mathbf{C}^{T})(\mathbf{X}^{T})\!\in\!\mathbb{R}^{L\times D'},\\
\mathbf{Y}^{F}&=\mathrm{SSD}(\mathbf{A}^{F},\mathbf{B}^{F},\mathbf{C}^{F})(\mathbf{X}^{F})\!\in\!\mathbb{R}^{L\times D'}.
\end{aligned}
\end{equation}

\begin{figure}[t!]
\begin{center}
\includegraphics[width=1\linewidth]{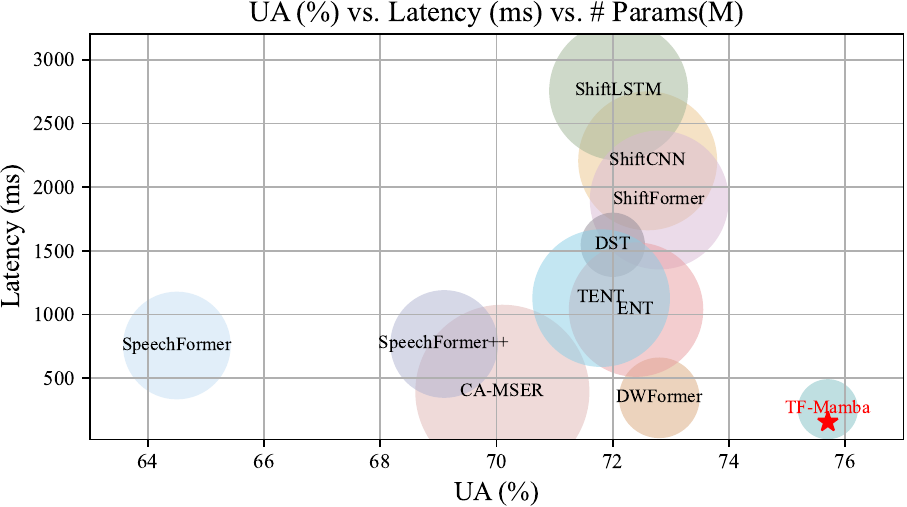}
\vspace{-2 em}
\caption{Our TF-Mamba achieves the SOTA performance on the SER task while being computationally efficient.}
\label{fig:param}
\end{center}
\vspace{-2 em}
\end{figure}

\subsection{Loss Function}
\subsubsection{\bfseries Complex Metric-Distance Triplet Loss}
Conventional Cross-Entropy Loss $\mathcal{L}_{\text{CE}}$~\cite{chen2023dwformer,li2021proposal} trains SER models by minimizing errors between predictions and true labels.
Still, it struggles to effectively capture subtle differences in the emotional feature space, particularly when the emotion classes are similar.
To address this, we propose adopting the existing InfoNCE triplet loss~\cite{wang2024frequency,zhou2022contrastive} to mitigate this shortcoming.
Given that speech signals contain rich acoustic nuances in the frequency domain, we extend the metric space to the complex frequency domain by FFT and construct a method using Vector Cosine Similarity $\mathrm{VecSim}(\cdot)$ to approximate the Euclidean distance, \ie, for any vectors $\vec U$ and $\vec V$, denoted as:
\begin{equation}
\begin{aligned}
\mathrm{VecSim}(\vec U,\vec V)=\frac{\mathbf{Re}(\vec U\cdot\overline{\vec V})}{|\vec U|\cdot|\vec V|},
\end{aligned}
\end{equation}
where $\mathbf{Re}(\cdot)$ and $|\cdot|$ are the real part and modulus operations respectively, and $\overline{\vec V}$ is the complex conjugate of $\vec V$.

Based on this, we build a CMDT loss $\mathcal{L}_c$ (illustrated in Fig.\ref{fig:overall}), which enhances the model's ability to distinguish between similar emotional features. This is achieved by simultaneously minimizing the distance between the anchor (predicted emotion complex vector $\{\vec{M}_{i}[k]\}$) and the positive anchor (true emotion complex vector $\{\vec{P}_{i}[k]\}$), while maximizing its distance from negative anchors (incorrect emotion complex vectors $\{\vec{G}_{j}[k]\}$). 
The loss is defined as:
{\small
\begin{equation}
\begin{aligned}
\mathcal{L}_{\text{CMDT}}=-\frac{1}{N}\sum_{i=1}^{N}\mathrm{log}\frac{\exp\left(\mathrm{VecSim}(\vec{M}_{i}[k],\vec{P}_{{i}}[k])/{\tau}\right)}{\sum_{j=1}^{N}\exp\left(\mathrm{VecSim}(\vec{M}_{i}[k],\vec{G}_{j}[k])/{\tau}\right)},
\end{aligned}
\end{equation}}
\noindent where $N$ means the number of samples in a batch, and $\tau$ is the temperature coefficient for regularizing the score distribution.

\subsubsection{\bfseries Optimization}
The overall loss function $\mathcal{L}_{\text{SER}}$ is the weighted sum of the following two loss terms:
\begin{equation}
\begin{aligned}
\mathcal{L}_{\text{SER}}=\mathcal{L}_{\text{CE}}+\lambda\mathcal{L}_{\text{CMDT}}.
\end{aligned}
\end{equation}

\section{EXPERIMENT}
\subsection{Experiment Setup}\label{DAT}
\subsubsection{\bfseries Datasets \& Evaluation Metrics}
Following the standard practice in SER~\cite{chen2023dwformer}, we conduct experiments on two benchmarks, \ie, IEMOCAP~\cite{busso2008iemocap} and MELD~\cite{meld}.
IEMOCAP includes 4 emotions (happy \& excited, angry, sad, and neutral), while MELD comprises 7 emotions (anger, disgust, fear, joy, neutral, sadness, and surprise). We report the performance on Weighted F1 (WF1\%), Weighted Accuracy (WA\%), and Unweighted Accuracy (UA\%) metrics. 
\subsubsection{\bfseries Implementation Details}
We standardize the raw audio signals to a 16 kHz sampling rate and use WavLM embeddings~\cite{chen2022wavlm} with a dimension of 1024, coupled with the self-attention encoder comprising 8 heads. 
To ensure fair comparisons, we follow the settings in \cite{chen2023dwformer,chen2023dst,shen2023mingling}, using five-fold cross-validation exclusively for the IEMOCAP dataset with a fixed batch size of 32. We employ the AdamW optimizer with an initial learning rate of 5$\times 10^{-4}$. In addition, the weight coefficient $\lambda$ in the loss function is set to 0.1.

\subsection{Results and Comparison}\label{RAC}

\begin{table}[t!]
\caption{~Performance comparison with the SOTA methods on the IEMOCAP and MELD datasets, where $^\ast$ denotes evaluation results from the published paper.}
\vspace{-2.2 em}
\begin{center}
\resizebox{1 \linewidth}{!}{
\begin{tabular}{l|c|c|c|c|c}
\hline
\multirow{2}{*}{Methods} & \multirow{2}{*}{Venue} & \multirow{2}{*}{Modality Type} & \multicolumn{2}{c|}{IEMOCAP} & MELD  \\
\cline{4-5}\cline{5-5}\cline{6-6}
&  & & WA(\%) $\uparrow$ & UA(\%) $\uparrow$ & WF1(\%) $\uparrow$	 \\
\hline
CA-MSER~\cite{zou2022speech} & ICASSP'22 & Spec+Logmel+Speech Feature & 68.9 & 70.1 & 42.2  \\
SpeechFormer~\cite{chen2022speechformer} & InterSpeech'22 & Speech Feature & 62.9 & 64.5 & 41.9  \\
SpeechFormer++~\cite{chen2023speechformer++} & TASLP'23 & Speech Feature & 68.8 & 69.1 & 46.0  \\
ShiftCNN~\cite{shen2023mingling} & ICASSP'23 & Speech Feature & 71.9 & 72.6 & 43.1 \\
ShiftLSTM~\cite{shen2023mingling} & ICASSP'23 & Speech Feature & 71.4 & 72.1 & 45.0 \\
ShiftFormer~\cite{shen2023mingling} & ICASSP'23 & Speech Feature & \underline{72.1} & \underline{72.8} & 45.6  \\
DWFormer\cite{chen2023dwformer} & ICASSP'23 & Speech Feature & 71.8 & \underline{72.8} & \underline{46.7} \\
DST~\cite{chen2023dst} & ICASSP'23 & Speech Feature & 70.9 & 72.3 & 45.8 \\
MSTR$^\ast$~\cite{li2023multi} & InterSpeech'23 & Speech Feature & 70.6 & 71.6 & 46.2 \\
ENT~\cite{shen2024asr} & ICASSP'24 & Speech Feature & 71.6 & 72.4 & 46.3  \\
TENT~\cite{shen2024asr} & ICASSP'24 & Speech Feature & 71.4 & 71.8 & 45.1  \\ \hline
 \textbf{TF-Mamba (Ours)} &  -  & Speech Feature & \textbf{75.3} \textcolor{red}{\scriptsize ↑3.2\%} & \textbf{75.7} \textcolor{red}{\scriptsize ↑2.9\%} & \textbf{48.5} \textcolor{red}{\scriptsize ↑1.8\%} \\
\hline
\end{tabular}
}
\label{tab:model_comparison}
\vspace{-1.5 em}    
\end{center}

\end{table}

\begin{table}[t!]
\caption{~Comparison of different pre-trained language models used in TF-Mamba on the IEMOCAP and MELD datasets.}
\vspace{-0.7 em}
\centering
\setlength{\tabcolsep}{8pt}
\resizebox{1 \linewidth}{!}{
\begin{tabular}{l|c|cc|c}
\hline
\multirow{2}{*}{PTMs} & \multirow{2}{*}{Venue} & \multicolumn{2}{c|}{IEMOCAP} & MELD \\
\cline{3-5}
& & WA(\%) $\uparrow$ & UA(\%) $\uparrow$ & WF1(\%) $\uparrow$ \\
\hline
 Wav2vec2-base~\cite{baevski2020wav2vec} & NeurIPS'20 & 54.4 & 55.4 & 31.9 \\
 Wav2vec2-large~\cite{baevski2020wav2vec} & NeurIPS'20 & 58.4 & 59.4 & 35.2 \\
 HuBERT-base~\cite{hsu2021hubert} & TALSP'21 & 62.7 & 62.9 & 37.7 \\
 HuBERT-large~\cite{hsu2021hubert} & TALSP'21 & 67.7 & 68.7 & 42.9 \\
 WavLM-base~\cite{chen2022wavlm} & JSTSP'21 & 71.7 & 70.3 & 42.5 \\
\textbf{WavLM-large}~\cite{chen2022wavlm} & \textbf{JSTSP'21} & \textbf{75.3} & \textbf{75.7} & \textbf{48.5} \\
 Whisper-base~\cite{radford2023robust} & ICML'23 & 62.8 & 61.0 & 34.3 \\
 Whisper-large~\cite{radford2023robust} & ICML'23 & 68.4 & 66.7 & 40.2 \\
\hline
\end{tabular}
}
\label{tab:ptm}
\vspace{-1 em}
\end{table}

\subsubsection{\bfseries Comparison with SOTA Methods}
Table~\ref{tab:model_comparison} reports the comparison between TF-Mamba and existing SOTA methods. 
Under identical experimental settings, TF-Mamba significantly 
outperforms SOTA methods across both datasets. 

\subsubsection{\bfseries Computational Complexity Analysis}
In Fig.~\ref{fig:param}, we compare the model parameters (M) and computational latency (ms) across SOTA methods. The results demonstrate that the proposed TF-Mamba reduces the parameters by 1.81$\times$ and improves computational latency by 1.87$\times$ compared to the advanced ShiftFormer method~\cite{shen2023mingling}. These improvements highlight that TF-Mamba significantly lowers computational costs 
efficiency while maintaining high performance, making it more effective and practical for real-world applications.

\subsubsection{\bfseries Visualization of TF-Mamba}

\begin{figure}[t!]
\begin{center}
\includegraphics[width=1\linewidth]{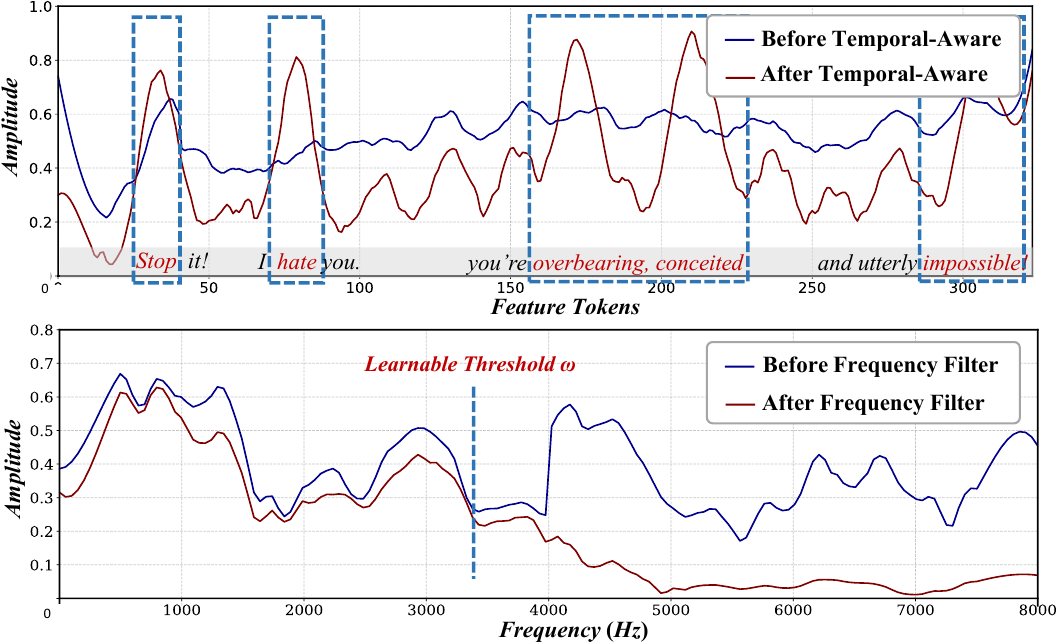}
\vspace{-1.7 em}
\caption{The top displays the feature token intensity comparison before and after the Temporal-Aware Module, while the bottom shows the spectrum comparison before and after the Frequency Filter Module.}
\label{fig:inflect}
\end{center}
\vspace{-1.5 em}
\end{figure}

Fig.~\ref{fig:inflect} visualizes the changes in speech features after processing through the temporal and frequency modules. 
In the temporal domain, emotionally salient features are emphasized while irrelevant features are down-weighted. In the frequency domain, high-frequency noise is reduced, minimizing interference in emotion detection.


\subsection{Ablation Study}\label{AS}
\subsubsection{\bfseries Effects of Pre-Trained Models}
To investigate the impact of different pre-trained models (PTMs) on SER accuracy, we compare various speech feature extractors. 
In Table~\ref{tab:ptm}, WavLM-Large significantly outperforms the other PTMs in SER performance.

\subsubsection{\bfseries Necessity of Key Components}
Table~\ref{tab:ab} presents the impact of each key component on performance. 
The baseline model only input speech embeddings into an emotion classifier for SER. 
Notably, to verify the necessity of introducing the Frequency-Domain Mamba, we compare TF-Mamba with the Dual Temporal-Domain Mamba. The results show that WA, UA, and WF1 have 2.9\%, 3.0\%, and 3.0\% gains on the IEMOCAP and MELD datasets, respectively, highlighting the effectiveness of the Frequency-Domain Mamba.


\begin{table}[t!]
\caption{~Ablation results of main components in TF-Mamba.}
\vspace{-0.7 em}
\centering
\resizebox{1.0\linewidth}{!}{
\begin{tabular}{l|c|cc|c}
\hline
\multirow{2}{*}{Networks} & \multirow{2}{*}{\#Params} & \multicolumn{2}{c|}{IEMOCAP} & MELD \\
\cline{3-5}
& & WA(\%) $\uparrow$ & UA(\%) $\uparrow$ & WF1(\%) $\uparrow$\\
\hline
Baseline & 0.66 M & 64.5 & 65.3 & 40.3 \\
\hline
+ Self-Attention Encoder & 5.25 M& 68.4 & 69.9 & 41.0 \\
\ + Temporal-Domain Mamba & 14.39 M & 71.7 & 72.3 & 44.3 \\
\ \ + Frequency-Domain Mamba & 20.79 M & 74.6 & 74.3 & 47.8 \\
\ \ \ \textbf{+ CMDT Loss (Full model)} & \textbf{20.79 M} & \textbf{75.3} &\textbf{75.7} &\textbf{48.5} \\ \hline
Dual Temporal-Domain Mamba & 20.77 M & 72.4 & 72.7 & 45.5 \\
\hline
\end{tabular}
}
\label{tab:ab}
\vspace{-1 em}
\end{table}

\section{Conclusions}
This paper proposes TF-Mamba, a novel multi-domain learning paradigm for fine-grained Speech Emotion Recognition (SER). By combining temporal awareness and frequency filtering, TF-Mamba provides comprehensive emotional representation with reduced computational costs. 
Additionally, the proposed CMDT loss enhances emotional discrimination. 
Experimental results show that TF-Mamba outperforms existing methods on multiple benchmarks, offering a more practical solution for future SER tasks. 


\vfill\pagebreak
\bibliographystyle{IEEEbib}
\bibliography{refs}

\end{document}